\documentclass[sigconf,nonacm]{acmart}

\AtBeginDocument{%
  \providecommand\BibTeX{{%
    \normalfont B\kern-0.5em{\scshape i\kern-0.25em b}\kern-0.8em\TeX}}}

\copyrightyear{2024}
\acmYear{2024}
\setcopyright{rightsretained}
\acmConference[CHI EA '24]{Extended Abstracts of the CHI Conference on Human Factors in Computing Systems}{May 11--16, 2024}{Honolulu, HI, USA}
\acmBooktitle{Extended Abstracts of the CHI Conference on Human Factors in Computing Systems (CHI EA '24), May 11--16, 2024, Honolulu, HI, USA}
\acmDOI{10.1145/3613905.3643981}
\acmISBN{979-8-4007-0331-7/24/05}

\begin{document}

\title[DCitizens Roles Unveiled]{DCitizens Roles Unveiled: SIG Navigating Identities in Digital Civics and the Spectrum of Societal Impact}

\author{Anna R. L. Carter}
\affiliation{%
 \institution{Northumbria University}
 \city{Newcastle}
 \country{UK}}
 
\author{Kyle Montague}
\affiliation{%
 \institution{Northumbria University}
 \city{Newcastle}
 \country{UK}}
 
\author{Reem Talhouk}
\affiliation{%
 \institution{Northumbria University}
 \city{Newcastle}
 \country{UK}}
 
\author{Shaun Lawson}
\affiliation{%
 \institution{Northumbria University}
 \city{Newcastle}
 \country{UK}}

\author{Hugo Nicolau}
\affiliation{%
 \institution{ITI/LARSyS, Instituto Superior Técnico, Universidade de Lisboa}
 \city{Lisbon}
 \country{Portugal}}
 
\author{Ana Cristina Pires}
\affiliation{%
 \institution{ITI/LARSyS, Instituto Superior Técnico, Universidade de Lisboa}
 \city{Lisbon}
 \country{Portugal}}
 
\author{Markus Rohde}
\affiliation{%
 \institution{University of Siegen}
 \city{Siegen}
 \country{Germany}}
 
\author{Alessio Del Bue}
\affiliation{%
 \institution{Istituto Italiano Di Technologia}
 \city{Genova}
 \country{Italy}}

\author{Tiffany Knearem}
\affiliation{%
 \institution{Google}
 \city{Boston}
 \country{USA}}

\renewcommand{\shortauthors}{Carter et al.}

\begin{abstract}
The DCitizens SIG aims to navigate ethical dimensions in forthcoming Digital Civics projects, ensuring enduring benefits and community resilience. Additionally, it seeks to shape the future landscape of digital civics for ethical and sustainable interventions. As we dive into these interactive processes, a challenge arises of discerning authentic intentions and validating perspectives. This exploration extends to evaluating the sustainability of future interactions and scrutinising biases impacting engaged communities. The commitment is to ensure future outcomes align with genuine community needs and address the ethical imperative of a considerate departure strategy. This dialogue encourages future researchers and practitioners to integrate ethical considerations and community-centric principles, fostering a more sustainable and responsible approach to technology-driven interventions in future urban regeneration and beyond.
\end{abstract}



\begin{teaserfigure}
\centering
    \includegraphics[width=\textwidth]{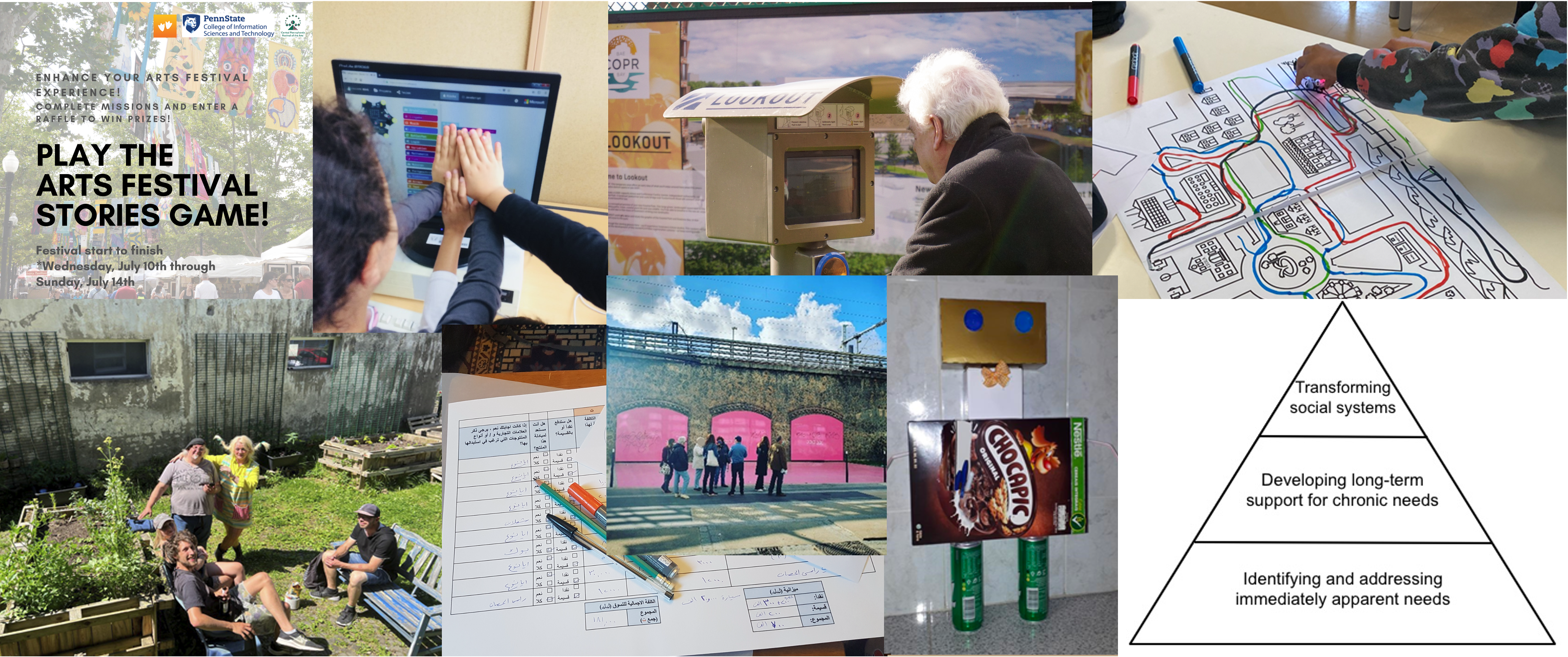}
    \caption{A montage of digital civics work completed by the authors. From top left to bottom right: 
    1) An arts festival scavenger hunt mobile app study on community engagement and reflection \cite{knearem21art}; 
    2) Come In Computer Clubs in Siegen \cite{aal18comp}; 
    3) A tangible embedded interface designed for the Covid Era to enhance topophilia in a changing city space \cite{carter22lookout, carter21masters}; 
    4) PartiPlay: A Participatory Game Design Kit for Neurodiverse Classrooms \cite{piedade23parti}; 
    5) Urban/Community Gardening in Siegen \cite{stickle14garden}; 
    6) Fieldwork surrounding food insecurity for refugees \cite{talhouk22foodinsec}; 
    7) MEMEX project,  a new open-source knowledge graph that facilitated assisted storytelling for Barcelona's migrant women \cite{nisi23connected};
    8) Community-Based Robot Design for Classrooms with Mixed Visual Abilities Children \cite{neto21robot};     
    9) Hierarchy of community needs pyramid \cite{knearem22covid}.}
    \Description{This figure is showing a montage of figures that relate to projects that authors on this SIG have completed. From top left to bottom right: 1) A project flyer which has a call for participation on inviting participants to attend an Arts Festival using an app with stories and games; 2) A student and researcher are interacting with a game on a computer as part of a computer club providing free computer use and training; 3) A participant interacts with the 'Lookout' a tangible embedded interface that enables participants to view the future of a city undergoing major regeneration; 4) A pupil is colouring in a map as part of a participatory game design for neurodiverse students; 5) Participants sit and chat within the community garden set up for the community in Siegen; 6) Tangible kit used within fieldwork projects about food security with refugees, including paper pads and a range of pens; 7) A robot built using a range of recycling products such as a cereal box and sprite cans as part of a community based project for children with mixed visual abilities; 8) A hierarchy pyramid created during a PhD Thesis focusing on the needs of the community within participatory design projects.}
    \label{fig:teaser}
\end{teaserfigure}

\maketitle

\section{Motivation and Background}
Digital Civics is a cross-disciplinary field that advocates the use of technology to empower citizens and non-state actors to cocreate, take an active role in shaping agendas, make decisions about service provision, and make such provisions sustainable and resilient \cite{vlachokyriakos16dc}. In particular, our focus here is on how digital technologies can encourage the move from transactional to relational service models and the potential of such models to reconfigure power relations between citizens, communities, and institutions \cite{olivier15digciv}. Digital civics aims to take an inclusive, participatory approach to the design and evaluation of new technologies and services that support ‘smart'; ‘data-rich' living across a range of communities.

The search for ‘Digital Civics' in the ACM DL for CHI uncovered $190$ papers that explore a range of projects within the realm of digital civics. Since the first paper within this field to be presented at CHI in $1996$ (\cite{Damer96first}), the number of papers has steadily risen, with a surge of over $500$\% observed from the $2010$s to the $2020$s, indicating a continuous acceleration in research output. Whilst there was a SIG on the subject of digital civics in $2016$ there have been $120$ papers presented at CHI alone since then, with surely many more across other conferences. This consistent pattern highlights the growing significance and sustained interest in digital civics, presenting new challenges and methods for discussion to shape this expanding field.


Digital Civics is a broad umbrella term that includes a range of technological interventions from games \cite{lazem17play,hitron18outdoor} to tangible interfaces \cite{dylan21lanterns,carter22lookout}, and storytelling applications \cite{halpering23probing,bidwell10mobile}, to name a few. As well as diverse interaction methods projects have spanned various themes, with activism \cite{bardzell11sex,michie18her}, health \cite{rooksby19student,singh19emot}, well-being \cite{oguamanam23intersect,zhu17digital}, gender \cite{mustaga19finance,tseng22care}, and safety \cite{freed23safety,chordia23deceptive} among some of the most common to be explored. Whilst the HCI field often works across a wide variety of communities we believe an intriguing aspect of digital civics research is its focus on providing spaces and opportunities for those less commonly considered within community landscapes. For example, studying the experiences of refugees \cite{talhouk16refugees,jensen20civic}, children \cite{vered98school,lu11creative}, children with disabilities \cite{PIRES2022iceta, tactopi, neto21robot}, grassroots initiatives \cite{green15beyond,rifat22social}, and the challenges faced in the developing world \cite{kam09game,frohlich09story}. Whilst the intricate aspects or backgrounds of these projects can be focused on certain themes, the overarching ethos and methods are broadly similar. This multifaceted exploration reflects the intricate nature of digital civics research, capturing a comprehensive and nuanced understanding of the interplay between societal issues, diverse communities, and cutting-edge technological advancements.

While the aforementioned papers focus on CHI, our considerations extend towards the broader community (i.e., beyond CHI and HCI), given that digital civics spans numerous domains. Looking further at the papers submitted within the wider ACM community, a total of $119$ papers have been published since $2015$ (amounting to $309$ across the ACM), with the majority appearing in Designing Interactive Systems (DIS: $22$), Computer-Supported Cooperative Work (CSCW: $27$) and Interactions ($17$). However, an exploration into alternative domains reveals a limited amount of research. Despite strides in philosophy, social science, and education-based research since $2014$, they have not expanded at the same rate as the HCI field, with engineering, design and psychology at an even lower rate. Therefore, we do not only aim to foster a community of growth within HCI but also aim to establish community events and hubs, such as this SIG, to generate increased interest and support for digital civics across diverse disciplines. Given that this research transcends specific domains and embodies a universal concept, it presents a compelling prospect for a SIG and future conference possibilities.

The expanding scope of the field has prompted a series of inquiries into our perceived roles within digital civics. Against the backdrop of heightened discussions on participatory design and increased emphasis on engagement with local communities from diverse perspectives, the following questions emerge: What motivates our involvement? Personal gain, a desire for heroism, or a genuine commitment to societal betterment? Or is it a strategy to achieve elevated academic visibility and recognition? Within these dynamic spaces, we find ourselves assuming various roles, be it as researchers, agents of change, activists, software engineers, or catalysts for positive transformation. These roles may appear clear to us as individuals but are not so black and white and may be completely different from the perceptions of the communities we aim to help.

Furthermore, as we delve deeper into these interactive processes, a necessity arises to scrutinise the authenticity of our intentions. Distinguishing whose intentions hold validity and whose perspectives carry weight becomes a nuanced challenge. This exploration naturally extends to considerations of the sustainability of these interactions. It compels us to assess how biases, particularly those acquired to advance professional trajectories, impact the communities we engage with. Striking a balance between meeting the authentic needs of the community and fulfilling the academic requirements to advance in an academic career can pose a significant challenge. Equally pressing is the question of how we cautiously conclude our involvement with these communities once our objectives have been met, e.g., if funding for a project is only available for one year. Therefore, the ethical imperative goes beyond the traditional model of intervention, emphasising the importance of a thoughtful departure strategy. Discussing the sustainability of interactions prompts reflection on the long-term impact of our engagement, ensuring that the benefits are enduring and that the community is left in a position of strength rather than vulnerability. We hope this dialogue and discussion will encourage our attendees to incorporate ethical considerations and community-centred principles into their projects, fostering a more holistic and responsible approach to technology-driven interventions in urban regeneration and beyond.

In extending an invitation to fellow researchers and industry professionals in the realm of digital civics, we aspire to initiate a formal and comprehensive discourse on these topics. Our objective is to articulate perspectives, derive insights, and lay the groundwork for collaboratively creating a guiding framework that future participatory designers can draw upon when navigating the complexities of this domain.

\section{SIG Proposal and Goals}
The DCitizens SIG has been designed to foster an inclusive discussion on crucial aspects of digital civics. We aim to explore diverse perspectives on (1) the motivations behind our engagement in digital civics and the alignment of our roles with community perceptions, (2) the ethical considerations involved in entering and exiting digital civics projects to ensure lasting benefits and community strength, and (3) the key considerations shaping the future of digital civics for ethical and sustainable interventions.

The session aims to be an inclusive forum that will address a diverse array of concerns within the digital civics landscape. The session has been designed to cater to attendees with varying backgrounds and focus, ensuring that the discourse is accessible and relevant to all. The organisers belong to are from diverse institutions, industries and disciplines and, therefore, are well-equipped to accommodate and facilitate discussions across the theme of digital civics, fostering an environment where insights from different backgrounds converge and enrich the conversation. This session aims to brainstorm ideas, highlighting our dedication to creating a space that truly reflects the interdisciplinary nature of digital civics. The main goal of this SIG is to kick-start a community discussion surrounding digital civics and the roles we play, leading to an open and integrated group ready for collaboration. 

\section{Audience}
Building upon the foundations laid by our previous work \cite{talhouk22dial,talhouk20food,Puussaar22sense,wood18online, knearem22covid, knearem21food}, as well as the ongoing Horizon project \cite{dcitizensproj22} which brings together four institutions across Europe, as well as expanding our collaborations with industry and NGOs, we seek to attract researchers, practitioners, community members, and policymakers who share an interest in the human aspects of digital civics. DCitizens SIG also aims to foster greater discussion and networking. Within the field, bringing together diverse perspectives, i.e. What does digital civics mean across the globe in comparison to our local contexts?

\section{Format}
The format of the SIG will be interactive to foster deep discussions and enable networking. It will consist of round table discussions with post-it notes for brainstorming and idea generation, with a Miro board provided as a virtual alternative. Each table will be provided with a laptop to host online participants to ensure they are integrated with the in-person attendees. An appointed author for each table will facilitate smooth discussions, generate post-it notes for accessibility and promote ease of interaction. This format allows for both physical and virtual collaboration. The schedule is planned as follows:

\begin{enumerate}
    \item \textbf{Welcome (5 minutes):} Introduction of the DCitizens SIG background/goals followed by the formation of groups, $4-6$ attendees per group, depending on numbers. Each group will have an author to prompt discussion. 
    
    \item \textbf{Ice breaker (5 minutes):} Each attendee will introduce themselves to the group, including their name, institution and their research interests regarding digital civics.

    \item \textbf{Activity 1a (5 minutes):} The groups will discuss the motivations behind their engagement in digital civics. They will create post-it notes of each \emph{role} mentioned, e.g., researcher, activist, etc. This will create an overall picture of the perception of the roles of academic research across digital civics. 
    
    \item \textbf{Activity 1b (5 minutes):} The groups will discuss how these roles align with community perceptions. How do participants/NGOs/stakeholders perceive us? They will create post-it notes of each possible perceived \emph{role} mentioned. e.g., software engineer, etc. This will create an overall picture of the perception of academic research across communities.   
    
    \item \textbf{Discussion 1 (5 minutes):} The groups will discuss the comparisons and similarities between these two role types (i.e., Personal and Community perceived roles). What are the possibilities and challenges for these role perceptions within digital civics of the future?
    
    \item \textbf{Activity 2a (5 minutes):} The groups will discuss the ethical considerations they undertake when entering their projects in digital civics, in particular, related to any steps taken for expectation management for the exiting of projects. They will create post-it notes of each \emph{step, challenge} mentioned. This will create an overall picture of the ethical considerations to be considered when entering digital civics projects.     

    \item \textbf{Activity 2b (5 minutes):} The groups will discuss the ethical considerations they undertake when exiting their projects in digital civics; it is often difficult to pass over a project whilst ensuring lasting benefits for the community. Groups will create post-it notes for the \emph{steps, suggestions and challenges} faced when exiting projects. This will create an overall picture of the ethical considerations to be considered when exiting digital civics projects.

    \item \textbf{Discussion 2 (5 minutes):} The groups will discuss the connections between these two project stages and possible steps that could be taken at the beginning of projects to aid the sustainable withdrawal from communities at the end of the project.
    
    \item \textbf{Digital Civics Moving Forward (15 minutes):} Bringing together the previous discussions, groups will discuss their three key considerations they believe could shape the future of digital civics to create more ethical and sustainable interventions. They will write these onto three separate post-it notes ready for the next discussion.
    
    \item \textbf{Collaborating DCitizens (15 minutes):} The groups will come back together into one, presenting their three considerations to the room and placing the post-it notes together onto one wall. We will then ask participants to move next to the post-it notes that they believe they can best help to achieve (with one group member assigned to also move the online participants via laptop), enabling participants to identify future collaborations for enhancing the sustainability and future of digital civics.

    \item \textbf{Staying Connected (5 minutes):} The SIG will end by introducing the \href{DCitizens.discord.com}{DCitizens community} channel,  \href{https://groups.google.com/u/1/g/dcitizens-seminar-series}{Google Group} and an invitation to attend and present at the DCitizens ongoing seminar series. Photographs of all the attendees' post-it notes will also be taken and uploaded onto an open access Miro board for the community. The in-person participants will also be invited to have dinner together to continue the discussions.
\end{enumerate}

\section{SIG Outcomes and Next Steps}
Prior initiatives concentrated on the development and deployment of digital tools, platforms, and processes within a Digital Civics research agenda. This session shifts the focus towards probing questions, exploring motivations, ethical considerations, and the future trajectory of digital civics. DCitizens SIG will build upon the Discord group and seminar series already in motion and serve as an extension, not only expanding the community but also with aspirations to set up a SIGCHI chapter. Additionally, we plan to organise follow-on workshops at conferences, including Participatory Design Conference (PDC) $2024$ and CHI $2025$. This initiative seeks to enhance collaboration and foster knowledge exchange within the digital civics community. Our goal is to kickstart a continuous dialogue, expressing varied perspectives, extracting insights, and laying the groundwork for collaboratively creating a guiding framework that future participatory designers can draw upon when navigating the complexities of this domain.

\begin{acks}
We thank our funding bodies at the European Commission (101079116 Fostering Digital Civics Research and Innovation in Lisbon), EPSRC (EP/T022582/1 Centre for Digital Citizens - Next Stage Digital Economy Centre) and the Portuguese Recovery and Resilience Program (PRR), IAPMEI/ANI/FCT under Agenda C645022399-00000057 (eGamesLab).
\end{acks}

\bibliographystyle{ACM-Reference-Format}
\bibliography{dcitizens}

\end{document}